\documentstyle[amstex,twocolumn,pra,aps,epsfig]{revtex}

\begin{document}
\draft
\title{Quantum Theory for Generation of Nonclassical Photon Pairs by a
Medium with Coherent Atomic Memory}
\author{Narek Sisakyan and Yuri Malakyan}
\address{Institute for Physical Research, Armenian National Academy of 
Sciences\\
Ashtarak-2, 378410, Armenia\\
(March 23, 2005)\\
\medskip}
\author{{\small 
\parbox{14.2cm}{\small \hspace*{3mm}
We present a fully quantum mechanical treatment of recent experiments on
creation of collective quantum memory and generation of non-classically
correlated photon pairs from an atomic ensemble via the protocol of Duan et
al. [Nature {\bf 414,} 413(2001)].\ The temporal evolution of photon
numbers, photon statistics and cross-correlation between the Stokes and
anti-Stokes fields is found by solving the equation of motion for atomic
spin-wave excitations. We consider a low-finesse cavity model with
collectively enhanced signal-to-noise ratio, which remains still
considerably large in the free-space limit. Our results describe
analytically the dependence of quantum correlations on spin decoherence time and
time-delay between the write and read lasers and reproduce the observed data
very well including the generated pulse shapes, strong violation of
Cauchy-Schwarz inequality and conditional generation of anti-Stokes
single-photon pulse. The theory we developed may serve as a basic 
approach for quantum description of storage and retrieval of
 quantum information, especially when the statistical properties 
of non-classical pulses are studied.
 \\ [3pt]PACS numbers: 42.50.Gy, 42.50.Dv, 03.67.-a,42.50.Ct }}}
\address{}
\maketitle

\narrowtext

\section{\protect\bigskip {\protect\normalsize INTRODUCTION}}

For transfer of quantum states between different nodes of quantum networks,
the obvious choice is to employ photons as the fast and robust carriers of
quantum information \cite{gis}. However, future development of quantum
communication intimately depends on successful attempts to reduce strongly
the photon losses at large distances that limit the range of application of
this technique. One possible way to cover large distances is using quantum
repeaters \cite{brieg}, which combine the teleportation of entangled states
as a channel of quantum information transfer (entanglement swapping) \cite%
{Zuk,enk} with local atomic memories for quantum information. Among the
various schemes for physical implementation of quantum repeaters and, hence,
for realization of scalable long-distance communication \cite{DuKim,Bln,duan}%
, one of the most promising approach is the protocol of Duan, Lukin, Cirac
and Zoller (DLCZ) \cite{duan}. This proposal is a probabilistic scheme that
relies upon entanglement between distant atomic ensembles, which is created
via successful detection of single photons emitted by initially
indistinguishable sources.

In this scheme, by utilizing simple linear optical operations, a quantum
repeater protocol with built-in atomic memory and entanglement purification
is implemented, which is robust against the realistic imperfections, such as
spontaneous loss and coupling inefficiency, that enables one to overcome
photon attenuation and to communicate quantum states over arbitrary long
distances with only polynomial costs. In contrast to proposed earlier more
complex protocols \cite{brieg,enk,cir}, the DLCZ scheme is attainable for
current experimental technologies. This significant advance has been reached
by exploiting the collective enhancement of atom-light interactions provided
by optically thick atomic ensembles due to many atoms constructive
interference effects \cite{kuz,fleisch,hald,phill,liu} resulting in
collectively enhanced signal-to-noise ratio.

The initial step toward realization of DLCZ protocol is generation of an
entanglement between remotely located atomic ensembles. To that end the
atomic ensembles with all atoms are prepared initially in the ground state $%
\mid 1$ $\rangle $ (Fig 1a) are illuminated by a short-pulse weak
Raman-pumping laser (referred to as a write laser), that induces a
Stokes-photon emission in the transition $\mid 3$ $\rangle $ $\rightarrow
\mid 2$ $\rangle $. Due to coherent coupling of different atoms to the
Stokes pulse propagating collinearly with the pumping laser, a collective
atomic mode is created in the form of long-lived spin-wave excitations. As a
result, a strong correlation between these atomic and Stokes-light modes is
produced, while other optical modes are weakly correlated with the
collective atomic state and contribute to noise. This significant
enhancement of signal-to-noise ratio allows one to generate a pure
entanglement between two distinct atomic ensembles by interfering and
detecting single Stokes photons emitted from them. The challenging task for
this scheme is to demonstrate that the collective atomic state can be
separately measured and, hence, the quantum correlation between the atomic
and signal-Stokes modes can be revealed. This can be done by first
converting the atomic spin excitation into state of the single-mode idler
(anti-Stokes) light by applying, after a controllable time delay (Fig.1b), a
second laser beam (read laser), and then detecting the idler light again
through a single photon detector.

Very recently this program has been successfully accomplished in a number of
experiments on generation of nonclassical photon pairs both in samples of
cold atoms\cite{kuzm,chou,pol} and in room-temperature atomic vapor cells 
\cite{wei,eis}. An atomic vapor cell has been used also in another
experiment \cite{van} demonstrating convincingly a strong correlation
between the Stokes and anti-Stokes pulses in the regime of \ large photon
number. Meanwhile, a theoretical description of these processes is absent
with the exception of some particular results obtained for only the
Stokes-photon emission \cite{duan} or on the basis of phenomenological
approach \cite{kuzm}. Since in the DLCZ scheme we deal only with spontaneous
emission, a fully quantum treatment is needed.

In this paper, we develop the quantum theory of creation of atomic memory
and generation of nonclassical photon pairs via the DLCZ protocol. In our
approach, to exploit the collective enhancement of the atom-light coupling,
we consider the interaction between atoms and forward-scattered Stokes and
anti-Stokes photons in low-finesse ring cavity \cite{roch} with an
independent photon emission by individual atoms. The correlation between
Stokes and anti-Stokes photons, separated by time interval considerably
larger than the excited atomic state lifetime, is established via temporal
dynamics of the collective atomic mode. The equation of motion for the
latter is derived by adiabatic elimination of cavity modes in the bad-cavity
limit. The probability of excitation by the write and read lasers is assumed
to be small and is treated in a perturbative way. Our results describe
analytically the evolution of quantum correlations, including their
dependence on coherent atomic memory lifetime and on time delay between the
write and read laser pulses.

For future technological developments, new sources for single-photons are
needed, which produce one photon on demand,\ at a specific time and not at
random. A remarkable feature of the DLCZ protocol is the ability to produce
not only single-photon pulses, but also any specific multiphoton state on
demand via conditional measurement on quantum systems of correlated photon
pairs. Compared to other systems of conditional generation of single
photons, such as atomic cascades \cite{claus,Grang} and parametric
down-conversion \cite{hong}, this scheme has an advantage of simultaneous
control over both the photon number and the spatio-temporal shape of
generated pulses that has been recently demonstrated in \cite{chou,pol,eis}.
Our theory reproduces the prominent features of observed results including
pulse shapes and conditional generation of anti-Stokes single-photon pulse,
as well as strong violation of the Cauchy-Schwarz inequality. We believe
that the calculations with this simple model may directly be applied to
realistic situation of free-space atomic cells. This expectation is
supported also by an important observation \cite{mduan} that the
three-dimensional theory confirms the large enhancement of signal-to-noise
ratio predicted by simple cavity-QED models \cite{duan,hald,LukY}.

The outline of the paper is as follows. In section II we derive the master
equation for the collective atomic mode by eliminating the degrees of
freedom associated with the cavity fields. The mean numbers of the Stokes
and anti-Stokes photons inside and emitted from cavity are found in section
III. Here we compare our results for photon fluxes with the experimentally
observed pulse shapes. In section IV we calculate the photon statistics and
cross-correlation between the fields and show a strong violation of
Cauchy-Schwarz inequality. In section V we obtain the probability of
conditional generation of single anti-Stokes photon and show its dependence
on spin-decoherence time. The conclusions are summarized in section VI.

\section{\protect\normalsize DYNAMICAL EQUATION FOR COLLECTIVE ATOMIC MODE}

We consider a large ensemble of $N$ atoms with a level structure shown in
Fig.1 and assume that initially all the atoms are prepared by optical
pumping in the ground state $\mid 1$ $\rangle $. The sample in the state $%
\Psi _{0}$ (Fig.1b) interacts with a weak write pulse with the duration $%
T_{W}$. The write laser beam acts on the transition $\mid 1$ $\rangle $ $%
\rightarrow \mid 3$ $\rangle $ with a large detuning $\Delta _{W}\ $and Rabi
frequency $\Omega _{W}$ and induces spontaneous emission of a Stokes photon
in transition $\mid 3$ $\rangle $ $\rightarrow \mid 2$ $\rangle $ while
flipping an atomic spin into the second ground state $\mid 2$ $\rangle $.
Detection of a forward propagating Stokes photon projects the state of
atomic sample onto the nonclassical collective state $\Psi _{1}$ with the
symmetric distribution of the flipped spin excitations (see below). The read
pulse acting on the transition $\mid 2$ $\rangle $ $\rightarrow \mid 4$ $%
\rangle $ with the\ Rabi frequency $\Omega _{R}$ and duration $T_{R}$ is
applied after time delay $\tau _{d}$ and converts the stored spin
excitations into the anti-Stokes light.\ \ \ \ \ \ \ \ \ \ \ \ \ \ \ \ \ \ \
\ \ \ \ \ \ \ \ \ 

\noindent Different schemes can be used for this retrieval process. It is
possible to convert the stored spin excitations into an anti-Stokes photon
in the electromagnetically induced transparency (EIT \cite{harr})
configuration similar to that exploited in previous experiments \cite%
{phill,liu} for restoring the classical light pulse. This mechanism has been
used in the most of the experiments with DLCZ scheme \cite{kuzm,chou,pol,eis}%
. Another way to transfer efficiently a quantum state from atoms to light is
off-resonant Raman configuration demonstrated experimentally in \cite{wei}.
Bellow, we consider for simplicity the Raman configuration for emission of
anti-Stokes photon. Nevertheless, our results describe very well the
experimental data obtained with EIT configuration \cite{chou,eis}.

The fields of write and read lasers propagating along the axis of
pencil-shape atomic ensemble in z direction are given by$\ \ \ \ \ \ \ \ \ \
\ \ \ \ \ \ \ \ \ \ \ \ \ \ \ \ \ \ \ \ \ \ \ \ \ \ \ \ \ \ \ \ \ \ \ \ \ \
\ \ \ \ \ \ \ \ \ \ \ \ \ \ \ \ \ $%
\begin{equation}
E_{W,R}(t)=E_{W,R}f_{W,R}^{1/2}(t)\exp (ik_{W,R}z-i\omega _{W,R}t),
\label{1}
\end{equation}
where \ $f_{W,R}(t)$ are the temporal profiles of the pulses.

The positive-frequency parts of forward propagating Stokes and anti-Stokes
cavity fields at frequencies $\omega _{1}$ and\ $\omega _{2}$ are expressed,
respectively, in terms of annihilation (creation) operators \ $%
a_{1}(a_{1}^{+})$\ \ and\ $a_{2}(a_{2}^{+})$ as follows:

\begin{equation}
E_{i}^{(+)}(z,t)=-i(\frac{2\pi \omega _{i}}{V})^{1/2}a_{i}\exp
(ik_{i}z-i\omega _{i}t),\ \ \ i=1,2\ \   \label{2}
\end{equation}
where $V$ \ is a quantization volume, which is assumed to be equal to the
volume of the medium.

Owing to the large detunings \ $\Delta _{1}=\omega _{31}-\omega _{W}$\ \ and
\ $\Delta _{2}=\omega _{42}-\omega _{R}$\ \ \ of the write and read lasers
from the respective transitions, we can adiabatically eliminate the upper
states $\mid 3\rangle $ \ and $\mid 4\rangle $, \ \ which are different in
general. Then, under the conditions $(k_{W}-k_{1})L\leq 1$\ \ \ and \ $%
(k_{R}-k_{2})L\leq 1$\ , where $L$ is the length of the atomic sample, the
interaction Hamiltonian for total system in the rotating frame has the form 
\begin{equation}
H=i\hbar 
\mathrel{\mathop{\stackrel{N}{\sum }}\limits_{i=1}}%
\left[ G(t)\sigma _{21}^{(i)}a_{1}^{+}-F(t)\sigma _{21}^{(i)}a_{2}\right]
+h.c.  \label{3}
\end{equation}
In Eq.(3) the summation is taken over all atoms, \ $\sigma _{\alpha \beta
}^{(i)}=\mid \alpha \rangle $ $_{i}$\ $\langle \beta \mid $\ \ is the atomic
spin-flip operator in the basis of the two ground states $\mid 1\rangle $\ \
and $\mid 2\rangle $ \ \ for the i-th atom and\ \ \ \ \ \ \ \ \ \ \ \ \ \ \
\ 
\begin{eqnarray}
\ \ G(t) &=&g_{S}\frac{\Omega _{W}}{\Delta _{W}}f_{W}^{1/2}(t),\   \label{4}
\\
F(t) &=&g_{AS}\frac{\Omega _{R}}{\Delta _{R}}f_{R}^{1/2}(t),\   \label{5}
\end{eqnarray}
where the Rabi frequencies of classical fields are $\Omega _{W}=\mu
_{31}E_{W}/\hbar $ \ and $\Omega _{R}=\mu _{42}E_{R}/\hbar $ and$\ \ \ \ \ \
\ \ \ \ \ \ \ \ \ \ \ \ \ \ \ \ \ \ \ \ \ \ \ \ \ $%
\begin{equation}
\ \ g_{S}=%
{2\pi \omega _{1} \overwithdelims() V}%
^{1/2}\mu _{32}\ \ \ \text{and \ \ \ }g_{AS}=%
{2\pi \omega _{2} \overwithdelims() V}%
^{1/2}\mu _{41}  \label{6}
\end{equation}
are the atom - quantized fields coupling constants, $\mu _{ij}$ is the
dipole matrix element of the transition $\mid i\rangle \rightarrow $\ $\mid
j\rangle $. Note that the Stark shifts of the upper levels$\mid 3\rangle $
and$\mid 4\rangle $ induced by the write and read lasers are included in the
frequencies of Stokes and \ anti-Stokes photons, so that \ $\omega
_{1}=\omega _{W}$\ $-\omega _{21}$\ \ \ and $\omega _{2}=\omega _{R}$\ $%
+\omega _{21}$.

In terms of the collective \noindent spin operators\ \ \ \ \ \ \ \ \ \ \ \ \
\ \ \ \ \ \ \ \ \ \ \ \ \ \ \ \ \ \ \ \ \ \ \ \ \ \ \ \ \ \ \ \ \ \ \ \ \ \
\ \ \ \ \ \ \ \ \ \ \ \ \ \ \ \ \ \ \ \ \ \ \ \ \ \ \ 
\begin{equation}
\ \ \ S^{+}=%
{1 \overwithdelims() \sqrt{N}}%
\mathrel{\mathop{\stackrel{N}{\sum }}\limits_{i=1}}%
\sigma _{21}^{(i)},\ \ \ \ \ \ \ \ \ \ \ S=(S^{+})^{+}\ \   \label{7}
\end{equation}%
the Hamiltonian $(3)$\ is written as 
\begin{equation}
H=i\hbar \sqrt{N}\left[ G(t)S^{+}a_{1}^{+}-F(t)S^{+}a_{2}\right] +h.c.\ \ \ 
\label{8}
\end{equation}%
When all atoms are prepared initially in level $\mid 1\rangle ,$ the states
that are coupled by Hamiltonian (8) are totally symmetric. Particularly, if
the atomic ensemble is initially in the state $\Psi _{0}=\mid
1_{1},1_{2},....1_{N}\rangle $, then upon emitting one Stokes photon it
settles down into symmetric state $\Psi _{1}=\frac{1}{\sqrt{N}}%
\mathrel{\mathop{\stackrel{N}{\sum }}\limits_{i=1}}%
\mid 1_{1},..2_{i},....1_{N}\rangle $ (see Fig.1)$.$

The system evolution is described by the master equation for the whole
density matrix $\rho $\ for the atoms and cavity modes \cite{gard}

\begin{equation}
\ \frac{d\rho }{dt}=-\frac{i}{\hbar }[H,\rho ]+(\frac{d\rho }{dt})_{rel}
\label{9}
\end{equation}
where the second term in right hand side (rhs) accounts for all relaxations
in the system. With use of the Liouville operator $L[\hat{O}]\rho =2\hat{O}%
\rho \hat{O}^{+}-(\hat{O}^{+}\hat{O}\rho +\rho \hat{O}^{+}\hat{O})$ it is
written\ in the form 
\[
\ (\frac{d\rho }{dt})_{rel}=\ 
\mathrel{\mathop{\stackrel{2}{\sum }}\limits_{i=1}}%
k_{i}L[a_{i}]\rho +\frac{\Gamma _{W}(t)}{2}%
\mathrel{\mathop{\stackrel{N}{\sum }}\limits_{i=1}}%
L[\sigma _{21}^{(i)}]\rho 
\]
\begin{equation}
\quad \quad +\frac{\Gamma _{R}(t)}{2}%
\mathrel{\mathop{\stackrel{N}{\sum }}\limits_{i=1}}%
L[\sigma _{12}^{(i)}]\rho -\gamma _{c}%
\mathrel{\mathop{\stackrel{N}{\sum }}\limits_{i=1}}%
I^{(i)}[\rho -\rho (0)].  \label{10}
\end{equation}
The first term in rhs of this equation, represents the cavity output at the
frequencies \ $\omega _{i}($ $i=1,2)$ with $2k_{i}\ $being the photon number
damping rate for i-th mode, while the second and third terms describe
spontaneous decay from the upper atomic states $\mid $3$\rangle $ and$\mid $
4$\rangle $ into the ground levels $\mid $2$\rangle $ \ and $\mid $1$\rangle 
$ \ resulting in optical pumping (OP) to these states during the interaction
with the write and read lasers, respectively. The rates of OP are

\begin{equation}
\Gamma _{W}(t)=\frac{\Omega _{W}^{2}}{\Delta _{W}^{2}}f_{W}(t)\gamma _{32},\
\ \ \Gamma _{R}(t)=\frac{\Omega _{R}^{2}}{\Delta _{R}^{2}}f_{R}(t)\gamma
_{41},  \label{11}
\end{equation}
where $\ \gamma _{ij}$ \ is \ a partial decay rate of upper level $i$ to the
state $j$.

We do not consider here two other channels $(\mid 3$ $\rangle $ $\rightarrow
\mid 1\rangle $ and$\mid 4$ $\rangle $ $\rightarrow \mid 2\rangle )$ \ of
spontaneous emission, because they do not change the atomic spin
distribution. The last term in rhs of Eq.(10) introduces into the model the
relaxation of the atomic ground-state coherence at the rate $\gamma _{c},$
which is supposed to be much smaller compared to the optical coherence
damping rate $\gamma $. In this term $I^{(i)}=\sigma _{11}^{(i)}+\sigma
_{22}^{(i)}\ $\ \ represents the unit matrix in the basis of atomic states.

Upon introducing the operators of the atomic populations of the ground
states $\mid 1\rangle $ and $\mid 2\rangle ,$ $\ S_{\alpha }=%
\mathrel{\mathop{\stackrel{N}{\sum }}\limits_{i=1}}%
\sigma _{\alpha \alpha }^{(i)}\ ,\ i=1,2,$ \ we obtain from Eq.(10) the
following equation for average number of the atoms in the state $\mid
2\rangle :$\ \ \ \ \ \ \ \ \ \ \ \ \ \ 

\begin{eqnarray}
\frac{d\left\langle S_{2}\right\rangle }{dt} &=&(\frac{d}{dt}%
+2k)(n_{1}^{(in)}-n_{2}^{(in)})+\Gamma _{1}(t)N  \nonumber \\
&&-[\gamma _{c}+\Gamma _{2}(t)]\left\langle S_{2}\right\rangle  \label{12}
\end{eqnarray}
where \ $n_{i}^{(in)}=\langle a_{i}^{+}a_{i}\rangle $ $,$ $i=1,2,$\ are the
mean photon numbers in, respectively, the Stokes and anti-Stokes modes
inside the cavity. To simplify, the decay constants of photon numbers, $%
k_{1} $\ and $k_{2},$ are taken both equal to $k$. Since the average number
of output photons $n_{i}^{(out)}$ is determined by the equations 
\begin{equation}
\frac{dn_{i}^{(out)}}{dt}=2kn_{i}^{(in)},\ i=1,2,  \label{13}
\end{equation}
we have

\begin{eqnarray}
\frac{d\left\langle S_{2}\right\rangle }{dt} &=&\frac{d}{dt}%
(n_{1}^{(tot)}-n_{2}^{(tot)})+\Gamma _{1}(t)N  \nonumber \\
&&-[\gamma _{c}+\Gamma _{2}(t)]\left\langle S_{2}\right\rangle ,  \label{14}
\end{eqnarray}
where \ $n_{i}^{(tot)}=n_{i}^{(in)}$\ $+n_{i}^{(out)}$\ \ is the total
number of photons in i-th mode.

The physical meaning of this equation is obvious showing that the atomic
population in the state $\mid 2\rangle $ is proportional to the photon
number difference in Stokes and anti-Stokes modes. Also, it increases due to
OP from the state $\mid 1\rangle $, when the atomic sample interacts with
the write pulse, and this is described by the term $\Gamma _{1}N$. Later on,
it decreases at the rate $\Gamma _{2}$\ proportional to the read pulse
intensity, as well as due to escape of the atoms from the laser beams area
at the rate $\gamma _{c}$.

We now eliminate the cavity fields adiabatically treating the two-photon
interaction terms G(t) and F(t) in the Hamiltonian (8) in a perturbative
way. By adiabatic elimination we obtain the equation of motion for the
reduced density matrix of the atoms\ $\rho _{a}=Tr_{c}\rho \ $\ in the form\
\ \ \ \ \ \ \ \ \ \ \ \ \ \ \ \ \ \ \ \ \ \ \ \ \ \ \ \ \ \ \ \ \ \ \ \ \ \
\ \ \ \ \ \ \ \ \ \ \ \ \ \ \ \ \ \ \ \ \ \ \ \ \ \ \ \ \ \ \ \ \ \ \ \ \ \
\ \ \ \ \ \ \ \ \ \ \ \ \ \ \ \ \ \ \ \ \ \ \ \ \ \ \ \ \ \ \ \ \ \ \ \ \ \
\ \ \ \ \ \ 
\begin{equation}
\ \frac{d\rho _{a}}{dt}=\frac{\alpha (t)}{2}L[S^{+}]\rho _{a}+\frac{\beta (t)%
}{2}L[S]\rho _{a}+(\frac{d\rho _{a}}{dt})_{rel},\   \label{15}
\end{equation}
where 
\[
(\frac{d\rho _{a}}{dt})_{rel}=\frac{\Gamma _{W}(t)}{2}%
\mathrel{\mathop{\stackrel{N}{\sum }}\limits_{i=1}}%
L[\sigma _{21}^{(i)}]\rho _{a} 
\]
$\ \ \ \ $\ \ \ \ \ \ \ \ \ \ \ \ \ \ \ \ \ \ \ \ \ \ \ \ \ \ \ \ \ \ \ \ \
\ \ \ 
\begin{equation}
\quad \quad +\frac{\Gamma _{R}(t)}{2}%
\mathrel{\mathop{\stackrel{N}{\sum }}\limits_{i=1}}%
L[\sigma _{12}^{(i)}]\rho _{a}-\gamma _{c}%
\mathrel{\mathop{\stackrel{N}{\sum }}\limits_{i=1}}%
I^{(i)}[\rho _{a}-\rho _{a}(0)]  \label{16}
\end{equation}

and\ 
\begin{eqnarray}
\alpha (t) &=&\frac{2N}{k}G^{2}(t)=\alpha f_{W}(t),  \label{17} \\
\beta (t) &=&\frac{2N}{k}F^{2}(t)=\beta f_{R}(t)  \label{18}
\end{eqnarray}
are the Stokes gain and anti-Stokes absorption (from the state $\mid
1\rangle $) coefficients, respectively.

This is the central equation of our paper. In deriving Eq.(15), we have
assumed only, apart from the weak interaction condition, that the write and
read pulses are not overlapped in time. Using Eq.(15), we study in next
sections the quantum dynamics of the system including the evolution of
collective atomic mode, photon statistics and nonclassical correlations
between the photons.

\section{\protect\normalsize MEAN PHOTON NUMBERS. PULSE SHAPES.}

In this section we apply Eq.(15) to find the mean photon numbers in \
forward-scattered Stokes and anti-Stokes modes inside the cavity and the
photon fluxes, given by Eq.(13), as a function of time, as well as the
number of spin-wave excitations in the collective atomic mode.

For the overall atomic population in the state $\mid 2\rangle $ we obtain
from Eq.(15)

\begin{eqnarray}
\frac{d\left\langle S_{2}\right\rangle }{dt} &=&\alpha
(t)[N_{sp}(t)+1]-\beta (t)N_{sp}(t)+\Gamma _{1}(t)N  \nonumber \\
&&-[\gamma _{c}+\Gamma _{2}(t)]\left\langle S_{2}\right\rangle ,  \label{19}
\end{eqnarray}
where $N_{sp}=\left\langle S^{+}S\right\rangle $ is a number of spin-wave
excitations or in other words, a number of the atoms with flipped spin,
which form the collective atomic mode. It obeys the equation 
\begin{equation}
\frac{dN_{sp}}{dt}=\alpha (t)[N_{sp}(t)+1]-\beta (t)N_{sp}(t)+\Gamma
_{1}(t)-\Gamma _{tot}  \label{20}
\end{equation}
where $\Gamma _{tot}=\gamma _{c}+\Gamma _{1}(t)+\Gamma _{2}(t).$

In obtaining Eqs.(19,20), the commutation relation \ 
\begin{equation}
\left\langle \lbrack S,S^{+}]\right\rangle \simeq 1  \label{21}
\end{equation}%
has been used, which follows from weak interaction condition implying that
upon interacting with the write laser, almost all the atoms are maintained
in the ground state $\mid 1\rangle .$

The only difference between the Eqs.(19,20) is due to the relaxation terms.
Consequently, the total number of the atoms $\left\langle S_{2}\right\rangle 
$ \ in the state $\mid 2\rangle $ is essentially larger than the number of
spin-wave excitations, because the former is basically determined by the
optical pumping $\Gamma _{1}N$ from the\ ground state $\mid 1\rangle $,
whereas the collective atomic mode is generated at the rate $\alpha
(t)<<\Gamma _{1}N$ as a result of coherent interaction with the write laser.
At the same time, owing to the fact that the spin coherence and, hence $%
N_{sp}$ , decays at the OP rate of individual atom, the signal-to-noise
ratio in Eq.(20) given by $\alpha /\Gamma _{1}\ $and $\beta /\Gamma _{2}\ $%
is greatly enhanced due to the large factor of the atom number $N$ \cite%
{duan} (see also bellow).

From the comparison of the Eqs.(14) \ and (19) in the time intervals \ $%
0\leq t\leq T_{W}$ and $T_{2}\leq t\leq T_{2}+T_{R},$ where $%
T_{2}=T_{W}+\tau _{d}$ is the instance of read pulse switching on, we
immediately find the equations for the mean photon numbers in cavity modes

\begin{eqnarray}
\frac{dn_{1}^{(in)}}{dt} &=&\alpha (t)[N_{sp}(t)+1]-2kn_{1}^{(in)},
\label{22} \\
\frac{dn_{2}^{(in)}}{dt} &=&\beta (t)N_{sp}(t)-2kn_{2}^{(in)},\ \ \ \ 
\end{eqnarray}%
It is useful at this point to consider numerical estimations. Reasonable
parameters are: light wavelength $\lambda =800nm,$ $\Omega _{W}\sim \gamma ,$
$\Omega _{R}\sim 10\gamma ,$ $\gamma \sim 10^{7}s^{-1},$\ $\gamma _{c}\sim
10^{4}s^{-1},$ $L=1\div 10cm,$ $V\sim 1cm^{3},$ and $N\sim 10^{12}.$ In the
free-space limit, $k=c/L\sim 3\cdot 10^{9}s^{-1}$ is the inverse of the
propagation time of the pulses through the atomic sample. Then, from
Eqs.(17,18) one has $\alpha \sim 10^{6}s^{-1},$ $\beta \sim 10^{8}s^{-1},$ \
and \ $\Gamma _{1}\sim 10^{-5}\gamma ,$ $\Gamma _{2}\sim 10^{-3}\gamma \sim
\gamma _{c}.$

Thus, the signal-to-noise ratio ($\alpha /\Gamma _{1}\ $or $\beta /\Gamma
_{2}$) is about $\sim 10^{4}$. At the same time $\alpha ,\beta <<k$, so that
from Eqs.(22,23) at $kt>>1$ we readily find

\begin{eqnarray}
\ n_{1}^{(in)} &=&\frac{1}{2k}\alpha (t)[N_{sp}(t)+1],\   \label{24} \\
\ n_{2}^{(in)} &=&\frac{1}{2k}\beta (t)N_{sp}(t).\   \label{25}
\end{eqnarray}
From the Eqs.(13) the photon fluxes are given by

\begin{eqnarray}
\frac{dn_{1}^{(out)}}{dt} &=&\alpha (t)[N_{sp}(t)+1],  \label{26} \\
\frac{dn_{2}^{(out)}}{dt} &=&\beta (t)N_{sp}(t).  \label{27}
\end{eqnarray}
The solution of Eq.(20) with the initial value $N_{sp}(t=0)=0$\ has the form

\begin{equation}
\bigskip N_{sp}(t)=%
\mathrel{\mathop{\stackrel{t}{\int }}\limits_{0}}%
dt^{\prime }\alpha (t^{\prime })\exp \{%
\mathrel{\mathop{\stackrel{t}{\int }}\limits_{t^{\prime }}}%
d\tau \lbrack \alpha (\tau )-\beta (\tau )-\Gamma _{tot}(\tau )]\}\ \ 
\label{28}
\end{equation}

It is worth noting that if one neglects the relaxations terms in Eq.(20), $%
n_{1}^{(out)}(t)$\ and $N_{sp}(t)$\ obey the same equation during the write
pulse. This means that there is an unambiguous correspondence between the
number of detected Stokes photons and spin excitations stored in collective
atomic mode. In particular, upon interacting with the write laser 
\begin{equation}
n_{1}^{(out)}(T_{W})=N_{sp}(T_{W}).  \label{29}
\end{equation}
Simple expressions are found for rectangular laser pulses

\begin{equation}
n_{1}^{(in)}(t)=\frac{\alpha }{2k}e^{\alpha t},\ n_{1}^{(out)}(t)=e^{\alpha
t}-1,\ 0\leq t\leq T_{W}  \label{30}
\end{equation}
and

\begin{equation}
n_{2}^{(in)}(t)=\frac{\beta }{2k}n_{1}^{(out)}(T_{W})\ e^{-\beta
(t-T_{2})-\gamma _{c}t},  \label{31}
\end{equation}%
\[
n_{2}^{(out)}(t)=n_{1}^{(out)}(T_{W})\ e^{-\gamma _{c}t} 
\]%
\begin{equation}
\quad \quad \times \lbrack 1-e^{-\beta (t-T_{2})}]\ ,\ \ T_{2}\leq t\leq
T_{2}+T_{R}\   \label{32}
\end{equation}%
where $\alpha $ and $\beta $ are defined in Eqs.(17) and (18). It is\ seen
that always $\ n_{i}^{(in)}<<n_{i}^{(out)}$. From the Eqs.(30) and (32) it
follows also that the numbers of output photons in both modes are the same $%
n_{1}^{(out)}(T_{W})=n_{2}^{(out)}(T_{2}+T_{R})$ ( we replace hereafter, for
short, the time argument $T_{2}+T_{R}$ by $T_{R}$), provided that the
intensity of read laser is sufficiently large, $\beta T_{R}>>1,$\ and \ the
time delay between the laser pulses is shorter compared to the spin
decoherence time: $\gamma _{c}T_{2}\simeq \gamma _{c}\tau _{d}<<1.$

In Fig.2 we show the Stokes and anti-Stokes pulse shapes calculated by means
of the Eqs.(26,27) for the values of parameters given above and for almost
rectangular laser pulses with the rise and fall times much shorter than the
pulse duration. These results coincide with the theoretical calculations
presented in \cite{eis} and reproduce very well the experimental data
reported in \cite{pol,eis}, although in these experiments the anti-Stokes
pulse has been retrieved in EIT configuration. Fig.2a demonstrates the
transition from a spontaneous to stimulated emission of Stokes photons with
increasing of write pulse intensity. In Fig.2b, the flux of anti-Stokes
photons is depicted as a function of time for a fixed number of detected
Stokes photons $n_{1}^{(out)}(T_{W})=3$. The total number of emitted
anti-Stokes photons is determined by the areas of the corresponding peaks.
It can be shown that in the case of strong read pulse, the stored spin
excitations is completely converted into the anti-Stokes photons, i.e. $%
N_{sp}(T_{W})=n_{1}^{(out)}(T_{W})\simeq n_{2}^{(out)}(T_{R}).$ \ This is
evident also from Fig.3, where the evolution of spin-wave excitations is
shown for different values of read laser intensity with a fixed Raman
scattering rate $\alpha (t)$\ corresponding to $n_{1}^{(out)}(T_{W})=3$.
Indeed, as it follows from Fig.3, after interaction with the read laser, $%
N_{sp}(T_{R})=0$ for $\Omega _{R}>>\Omega _{W,}$ whereas a residual coherent
excitation is preserved in atomic ensemble, if $\ \Omega _{R}\simeq \Omega
_{W}$. To demonstrate how the spin decoherence deteriorates the atomic
memory, we show in Fig.3b the same calculations for the case of ten times
larger decoherence rate $\gamma _{c}$ and with the same time delay between
the write and read laser pulses. In this case the total \ number of
retrieved spin excitations and, hence, of produced anti-Stokes photons is
strongly reduced.

\section{\protect\normalsize PHOTON STATISTICS AND CORRELATIONS. VIOLATION
OF CAUCHY-SCHWARZ INEQUALITY.}

The nonclassical character of the Stokes and anti-Stokes fields generated in
the DLCZ scheme has been experimentally studied in Refs.\cite%
{kuzm,chou,pol,wei,eis} by observing the violation of the Cauchy-Schwarz
inequality.

It is well known \cite{man} that two electromagnetic fields, for which a
positive true probability distribution exists, satisfy the following
Cauchy-Schwarz inequality\ \ \ \ \ \ 
\begin{equation}
\left\vert g^{(12)}\right\vert ^{2}\leq g^{(11)}g^{(22)},\ \ \ \   \label{33}
\end{equation}%
where $g^{(ii)}$\ \ is the normalized second order auto-correlation
functions for i-th field and $g^{(12)}$\ is the cross-correlation between
the two fields. They are defined as\ 
\begin{equation}
g^{(ii)}(t)=\frac{G^{(ii)}(t)}{\left\langle n_{i}(t)\right\rangle ^{2}}=%
\frac{\left\langle a_{i}^{+}(t)a_{i}^{+}(t)a_{i}(t)a_{i}(t)\right\rangle }{%
\left\langle n_{i}(t)\right\rangle ^{2}},\quad i=1,2\   \label{34}
\end{equation}%
\begin{eqnarray}
\ \ g^{(12)}(t_{1},t_{2}) &=&\frac{G^{(12)}(t_{1},t_{2})}{\left\langle
n_{1}(t_{1})\right\rangle \left\langle n_{2}(t_{2})\right\rangle }  \nonumber
\\
&=&\frac{\left\langle
a_{1}^{+}(t_{1})a_{2}^{+}(t_{2})a_{2}(t_{2})a_{1}(t_{1})\right\rangle }{%
\left\langle n_{1}(t_{1})\right\rangle \left\langle
n_{2}(t_{2})\right\rangle }  \label{35}
\end{eqnarray}%
with\ \ $n_{i}(t)=\ \left\langle a_{i}^{+}(t)a_{i}(t)\right\rangle ,\ i=1,2$%
, being the mean photon numbers.

The inequality (33) is violated for quantized fields. In our case, the
correlation functions $G^{(ij)}(t)$ are easily calculated for the cavity
modes. Then, taking into account that in the bad-cavity limit the output
fields and cavity modes have obviously the same photon statistics, we apply
the obtained results for the detected photons.

Using the Hamiltonian (8), the Heisenberg-Langevin equations for the cavity
modes are given by\ \ \ \ \ \ \ \ \ \ \ \ \ \ \ \ \ \ \ \ 
\begin{eqnarray}
\ a_{1} &=&\sqrt{N}\frac{G(t)}{k}S^{+}+%
\mathrel{\mathop{\stackrel{t}{\text{ }\int }}\limits_{0}}%
dt^{\prime }F_{1}(t^{\prime })e^{-k(t-t^{\prime })},  \label{36} \\
\ \ a_{2} &=&\sqrt{N}\frac{F(t)}{k}S\ +%
\mathrel{\mathop{\stackrel{t}{\text{ }\int }}\limits_{0}}%
dt^{\prime }F_{2}(t^{\prime })e^{-k(t-t^{\prime })},\ \   \label{37}
\end{eqnarray}%
where the noise operators $F_{i}(t)\ $associated with cavity losses in the
Stokes and anti-Stokes modes have the properties \cite{scul} 
\begin{equation}
\langle F_{i}(t)\rangle =\langle F_{i}(t)F_{i}(t^{\prime })\rangle =\langle
F_{i}^{+}(t)F_{i}(t^{\prime })\rangle =0  \label{38}
\end{equation}%
\[
\langle F_{i}(t)F_{j}^{+}(t^{\prime })\rangle =2k_{i}\delta ^{ij}\delta
(t-t^{\prime }).
\]%
The Eqs.(36) and (37) reproduce the solutions (24) and (25) for mean photon
numbers $n_{1}^{(in)}$ and $n_{2}^{(in)}$, if the following conditions:\ \ \
\ \ \ \ \ \ \ \ \ \ \ \ \ \ \ \ \ \ \ \ 
\begin{equation}
\left\langle S(t)F_{i}(t^{\prime })\right\rangle =\left\langle
F_{i}^{+}(t^{\prime })S^{+}(t)\right\rangle =0,\ \ i=1,2  \label{39}
\end{equation}%
are satisfied for $t\geq $ $t^{\prime }.$ By using the solution for $S(t)$
obtained with the Hamiltonian (8) and applying again the properties of
Langevin forces (38) it may be proved directly that these correlations must
vanish. Moreover, the similar calculations show that this is true also in
general case of correlation functions with more than two field operators
written in normal order. This allows us, keeping only the first terms in
Eqs.(36) and (37), to express the correlation functions $G(t)$ in terms of
the atomic collective spin operators as \ $\ \ \ \ \ \ \ \ \ \ \ \ $%
\begin{eqnarray}
G^{(11)}(t) &=&\left( \frac{\alpha (t)}{2k}\right) ^{2}\Phi _{1}(t),\ \ \Phi
_{1}(t)=\left\langle S^{2}(t)S^{+2}(t)\right\rangle   \label{40} \\
G^{(22)}(t) &=&\left( \frac{\beta (t)}{2k}\right) ^{2}\Phi _{2}(t),\ \ \Phi
_{2}(t)=\left\langle S^{+2}(t)S^{2}(t)\right\rangle   \label{41}
\end{eqnarray}%
The equations for $\Phi _{1}(t)$\ \ and $\Phi _{2}(t)$\ are derived from the
master equation (15) 
\begin{eqnarray}
\frac{d\Phi _{1}(t\leq T_{W})}{dt} &=&[2\alpha (t)-\gamma _{c}]\ \Phi
_{1}(t),  \label{42} \\
\frac{d\Phi _{2}(t\geq T_{2})}{dt} &=&-[2\beta (t)+\gamma _{c}]\ \Phi
_{2}(t),  \label{43}
\end{eqnarray}%
and have the following simple solutions

\begin{equation}
\Phi _{1}(t)=\Phi _{1}(0)\exp \left\{ 
\mathrel{\mathop{\stackrel{t}{\int }}\limits_{0}}%
[2\alpha (t^{\prime })-\gamma _{c}]dt^{\prime }\right\} ,\text{ }t\leq T_{W}
\label{44}
\end{equation}
\begin{equation}
\Phi _{2}(t)=\Phi _{2}(T_{W})\exp \left\{ -%
\mathrel{\mathop{\stackrel{t}{\int }}\limits_{T_{W}}}%
[2\beta (t^{\prime })+\gamma _{c}]dt^{\prime }\right\} ,\text{ }t\geq T_{2}
\label{45}
\end{equation}
where the value $\Phi _{1}(0)=\left\langle S^{2}(0)S^{+2}(0)\right\rangle =2$%
\ \ is readily found recalling the commutation relation (21) and that $%
N_{sp}(0)=\ \left\langle S^{+}(0)S(0)\right\rangle =0$. To obtain $\Phi
_{2}(T_{W})$ we use the solution of the equation (28) at $t=T_{W}$\ 

\begin{equation}
N_{sp}(T_{W})=\left\langle S^{+}(T_{W})S(T_{W})\right\rangle =\exp [%
\mathrel{\mathop{\stackrel{T_{W}}{\int }}\limits_{0}}%
\alpha (z^{\prime })dz^{\prime }]-1  \label{46}
\end{equation}
which, together with \ $\Phi _{1}(T_{W})$\ from Eq.(44), yields

\begin{equation}
\Phi _{2}(T_{W})=\left\langle S^{+2}(T_{W})S^{2}(T_{W})\right\rangle
=2\left\langle S^{+}(T_{W})S(T_{W})\right\rangle ^{2}  \label{47}
\end{equation}
This relation indicates a chaotic nature of spin-wave bosonic excitations
just like a similar relation \ $\left\langle a^{+2}a^{2}\right\rangle =2$\ $%
\left\langle a^{+}a\right\rangle ^{2}$\ takes place for a thermal light.

By substituting the Eqs.(40,41,44,45,47) and (24,25) for mean photon numbers 
$n_{1}^{(in)}(T_{W})$ and \ $n_{2}^{(in)}(t)$ into Eqs.(34), we eventually
have

\begin{equation}
g^{(11)}(t\leq T_{W})=2\text{ \ and\ \ }g^{(22)}(t\geq T_{2})=2\exp [\gamma
_{c}(t-T_{W})]\text{\ }  \label{48}
\end{equation}
showing that both the Stokes and anti-Stokes modes satisfy Gaussian
statistics. It is worth noting that the correlations $g^{(ii)}$ are
independent of time apart from the last factor in $g^{(22)}(t)$\ \
indicating that an increase of time delay between the laser pulses or of
spin-decoherence rate $\gamma _{c}$ results in a superchaotic statistics of
the anti-Stokes field.

Similarly the cross-correlation function $G^{(12)}(t_{1},t_{2})$ \ may be
represented in the form

\begin{eqnarray}
G^{(12)}(t_{1},t_{2}) &=&\frac{\alpha (t_{1})}{2k}\frac{\beta (t_{2})}{2k}%
\Phi _{12}(t_{1},t_{2});\text{ \ }  \nonumber \\
\text{\ }\Phi _{12}(t_{1},t_{2}) &=&\left\langle
S(t_{1})S^{+}(t_{2})S(t_{2})S^{+}(t_{1})\right\rangle  \label{49}
\end{eqnarray}
where $t_{1}\leq T_{W}$ and \ $t_{2}\geq T_{2}$. The equation for $\Phi
_{12}(t_{1},t_{2})$ is deduced from Eq.(20) by applying the quantum
regression theorem \cite{gard,lax} with \ $t_{2}=$ $t_{1}+\tau ,$ where $%
\tau \geq \tau _{d},$\ \ yielding

\begin{equation}
\frac{d\Phi _{12}(t_{1},t_{1}+\tau )}{dt}=-[\beta (t_{1}+\tau )+\gamma
_{c}]\ \phi _{12}(t_{1},t_{1}+\tau );\text{ \ }  \label{50}
\end{equation}
the solution of which is

\begin{equation}
\Phi _{12}(t_{1},t_{1}+\tau )=\phi _{12}(t_{1},t_{1})\exp [%
\mathrel{\mathop{\stackrel{t_{1}+\tau }{-\int }}\limits_{t_{1}}}%
\beta (t^{\prime })dt^{\prime }-\gamma _{c}\tau ]  \label{51}
\end{equation}
Using Eq.(47), $\Phi _{12}(t_{1},t_{1})$ is easily transformed to

\begin{equation}
\Phi _{12}(t_{1},t_{1})=\left\langle S(t_{1})S^{+}(t_{1})\right\rangle \
[2\left\langle S^{+}(t_{1})S(t_{1})\right\rangle +1]  \label{52}
\end{equation}
From Eq.(33), the cross-correlation between the two modes is then obtained as

\begin{equation}
g^{(12)}(t_{1},t_{2})=2+\frac{1}{\left\langle
S^{+}(t_{1})S(t_{1})\right\rangle }  \label{53}
\end{equation}
and, thus, for the ratio between $\left| g^{(12)}\right| ^{2}$and $%
g^{(11)}\cdot g^{(22)}$\ at $t_{1}=T_{W}$ and $t_{2}=T_{2}+T_{R\text{ }}$ we
finally have

\begin{equation}
R=\frac{\left| g^{(12)}(t_{1},t_{2})\right| ^{2}}{g^{(11)}(t_{1})\
g^{(22)}(t_{2})}=\left| \frac{1+2n_{1}^{(out)}(T_{W})}{2n_{1}^{(out)}(T_{W})}%
\right| ^{2}\exp (-\gamma _{c}\tau _{d})  \label{54}
\end{equation}
where $\left\langle S^{+}(T_{W})S(T_{W})\right\rangle $ is replaced by the
number $n_{1}^{(out)}(T_{W})$ of \ Stokes photons emitted from cavity (see
Eq.(29)). Remind that in deriving Eqs.(48) and (54) we have neglected all
atomic relaxations during the interaction with laser pulses under
assumptions $\Gamma _{tot}\ll \alpha ,\beta $, but we have kept the term $%
\gamma _{c}\tau _{d}$, which is by no means small and leads to a loss of
atomic memory that is manifested by exponential decay of the ratio $R$.

It follows from Eq.(54) that if the write pulse is sufficiently weak, so
that the excitation probability $p$ in the Stokes channel is less than unity
and, hence, $n_{1}^{(out)}(T_{W})\simeq \alpha T_{W}=p\ll 1,$ the
Cauchy-Schwarz inequality is strongly violated by the law $R\sim 1/p^{2}\gg
1 $. In the experiments \cite{kuzm,chou,pol,wei} the violation of this
inequality has been tested to confirm the non-classical correlations between
the Stokes and anti-Stokes pulses and a significant violation has been
observed for small $p$, although, as has been discussed in \cite{kuzm,chou},
several imperfections limit, in practice, the degree of violation of
Cauchy-Schwarz inequality and lead to deviation from the ideal case
expressed by Eq.(54).

\section{\protect\normalsize PRODUCING SINGLE\ PHOTONS VIA CONDITIONAL
MEASUREMENT.}

The strong non-classical correlation between the Stokes and anti-Stokes
pulses in DLCZ scheme can be employed for producing single photons as a
result of conditional measurement on correlated photon pairs (see also \cite%
{Grang,hong,Pitt}), when a detection of one Stokes photon tightly projects
the anti-Stokes field into an one-photon state. In order to quantify the
confidence level of this procedure we analyze the third-order correlation
function \cite{man}

\begin{equation}
g^{(3)}(t_{1},t_{2},t_{2})=\frac{P_{c}(1_{2},1_{2}\mid 1_{1})}{\left\vert
P_{c}(1_{2}\mid 1_{1})\right\vert ^{2}},  \label{55}
\end{equation}%
which is less unity for quantized fields (it approaches zero in the case of
creation of single photons), while $g^{(3)}(t_{1},t_{2},t_{2})\geq 1 $ for
classical fields. In Eq.(55), $P_{c}(1_{2},1_{2}\mid 1_{1})$ and $%
P_{c}(1_{2}\mid 1_{1})$ are conditional probabilities for detection of two
and one photon from anti-Stokes field, respectively, conditioned upon the
detection of one photon in the\ Stokes channel. Using the known formulae 
\cite{man,lax}, the Eq.(55) may be transformed to

\begin{equation}
g^{(3)}(t_{1},t_{2},t_{2})=\frac{P(1_{1})P(1_{1},1_{2},1_{1})}{\left\vert
P(1_{1},1_{2})\right\vert ^{2}},  \label{56}
\end{equation}%
where the total probabilities are given by

\[
P(1_{1})=\left\langle a_{1}^{+}(t)a_{1}(t)\right\rangle =n_{1}^{(in)}(t_{1}) 
\]
\[
P(1_{1,}1_{2})=G^{(12)}(t_{1},t_{2})\approx \frac{\alpha (t_{1})}{2k}\frac{%
\beta (t_{2})}{2k}\ 
\]
\begin{equation}
\quad \times \left\langle S(t_{1})S^{+}(t_{1})\right\rangle \exp [%
\mathrel{\mathop{\stackrel{t_{2}}{-\int }}\limits_{t_{1}}}%
\beta (t^{\prime })dt^{\prime }-\gamma _{c}(t_{2}-t_{1)}]  \label{57}
\end{equation}
\begin{eqnarray}
P(1_{1},1_{2},1_{2}) &=&G^{(122)}(t_{1},t_{2},t_{2})  \nonumber \\
&=&\left\langle
a_{1}^{+}(t_{1})a_{2}^{+}(t_{2})a_{2}^{+}(t_{2})a_{2}(t_{2})a_{2}(t_{2})a_{1}(t_{1})\right\rangle
\nonumber
\end{eqnarray}
Here we have kept only the second term in $G^{(12)}(t_{1},t_{2})$ (see
Eqs.(49) and (52)) taking into account the smallness of \ $%
n_{1}^{(out)}(t_{1})=\left\langle S^{+}(t_{1})S(t_{1})\right\rangle $.

The last correlation function in Eqs.(57) is calculated by the same method
employed in previous sections

\begin{eqnarray}
G^{(122)}(t_{1},t_{2},t_{2}) &=&\frac{\alpha (t_{1})}{2k}\left( \frac{\beta
(t_{2})}{2k}\right) ^{2}\Phi (t_{1},t_{2},t_{2}),\ \   \nonumber \\
\ \Phi (t_{1},t_{2},t_{2}) &=&\left\langle
S(t_{1})S^{+2}(t_{2})S^{2}(t_{2})S^{+}(t_{1})\right\rangle ,\text{ \ }
\label{58}
\end{eqnarray}
where $t_{1}\leq T_{W},$ $t_{2}\geq T_{2}.$ We obtain the equation for $\Phi
(t_{1},t_{2},t_{2})$ by applying again the quantum regression theorem to the
Eq.(43) and find the following solution

\[
\Phi (t_{1},t_{2},t_{2})=\Phi (t_{1},t_{1},t_{1})\exp \left\{ -%
\mathrel{\mathop{\stackrel{t_{2}}{\int }}\limits_{t_{1}}}%
[2\beta (t^{\prime })+\gamma _{c}]dt^{\prime }\right\} 
\]%
where, for $\Phi (t_{1},t_{1},t_{1}),$ by solving the corresponding equation
obtained from Eq.(15), we have

\begin{equation}
\Phi (t_{1},t_{1},t_{1})=4[n_{1}^{(out)}(t_{1})+1]\left[ \frac{3}{2}%
n_{1}^{(out)}(t_{1})+1\right] n_{1}^{(out)}(t_{1})  \label{59}
\end{equation}%
Substituting Eqs.(57) and (58) into Eq.(56), we finally get

\begin{eqnarray}
g^{(3)}(T_{W},T_{R},T_{R}) &=&4n_{1}^{(out)}(T_{W})  \nonumber \\
&&\ \times \left[ \frac{3}{2}n_{1}^{(out)}(t_{W})+1\right] \exp (\gamma
_{c}\tau _{d})  \label{60}
\end{eqnarray}%
In the regime of very weak Raman excitation, $n_{1}^{(out)}(T_{W})\simeq
p\ll 1$, and provided that $\gamma _{c}\tau _{d}<1$ one has $g^{(3)}\simeq
4p\ll 1$ that corresponds to ideal case of single photon creation.

Experimentally, the conditional generation of single photons in the
anti-Stokes pulse has been recently demonstrated with $g^{(3)}\simeq 0.3$ in
an ensemble of cold $Cs$ atoms \cite{chou}. The possibility of conditional
preparation of Fock states with a given number of photons in optically dence
medium of $Rb$ atoms has been reported in Ref. \cite{eis}.\bigskip\ 

\section{\protect\normalsize SUMMARY}

In summary, we have developed a theory in the weak pumping limit to describe
analytically the non-classical correlations between the photons emitted
separately from a medium with macroscopic atomic memory for single-photon
fields. First, the spin excitations in atoms are produced in a Raman process
induced by the write pulse and are entangled with forward-scattered Stokes
photons, and, finally, they are converted by the read pulse into the
anti-Stokes photons with an efficiency that depends on the read laser
intensity. An important result of our study is that the correlations between
the photons are purely determined by only the atomic spin correlations. This
allows one to describe the evolution of the system by the dynamical equation
for the atoms, instead of solving the underlying equations for quantized
pulses. We have derived the master equation for atomic dynamics and have
found its analytical solutions for consecutive write and read pulses with
time separation longer than the duration of the pulses. The time dependence
of photon correlations shows that all quantum effects (violation of
Cauchy-Schwarz inequality, conditional generation of single photons, etc)
are robustly manifested within the atomic memory lifetime. Otherwise, they
exponentially disappear at the atomic spin decoherence rate. To minimize the
dissipation we considered the Raman scheme for mapping the atomic spin state
onto the anti-Stokes light. Our results obtained in bad-cavity limit are
consistent with those observed in single pass scheme with dense atomic
media, where for retrieving of the anti-Stokes light the EIT configuration
has been used \cite{chou,pol,eis}. From this point of view, the theory we
developed may serve as a basic approach for quantum description of storage
and retrieval of quantum information, especially when the statistical
properties of non-classical pulses are studied.

\bigskip

This work has been supported by the ISTC Grant No. A-1095.

\bigskip

\bigskip {\normalsize FIGURE CAPTIONS}

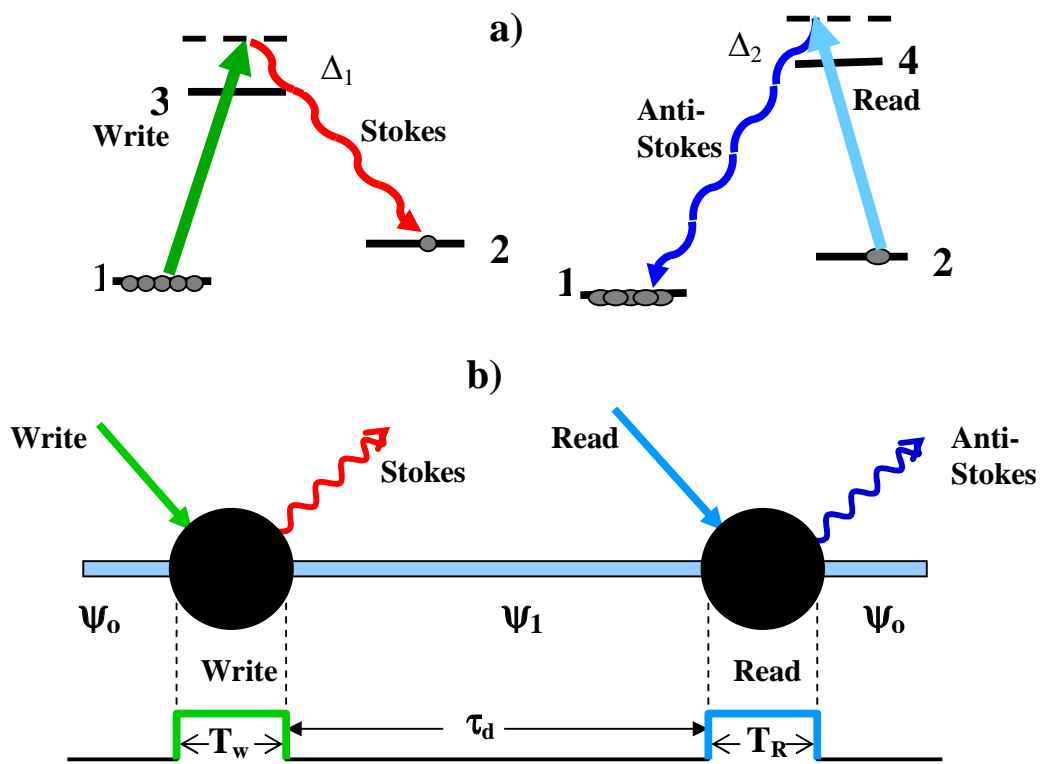
Fig.1. a) Atomic level structure for emission of Stokes and anti-Stokes
photons in off-resonant Raman configuration with one photon detunings $%
\Delta _{1}$ and $\Delta _{2},$ respectively. b) Interaction diagram in the
case of one Stokes-photon emmission, where in initial state $\Psi _{0}$ all
atoms are prepared in the ground state $\mid 1>.$ In \ intermediate stage,
the atomic ensemble is in a collective state $\Psi _{1}$ with symmetric
distribution of one atom excitation$.$ c) Time sequence for the write and
read pulses having durations T$_{W}$ and T$_{R}.$ $\tau _{d}$ is time-delay
between the pulses. \ \ 

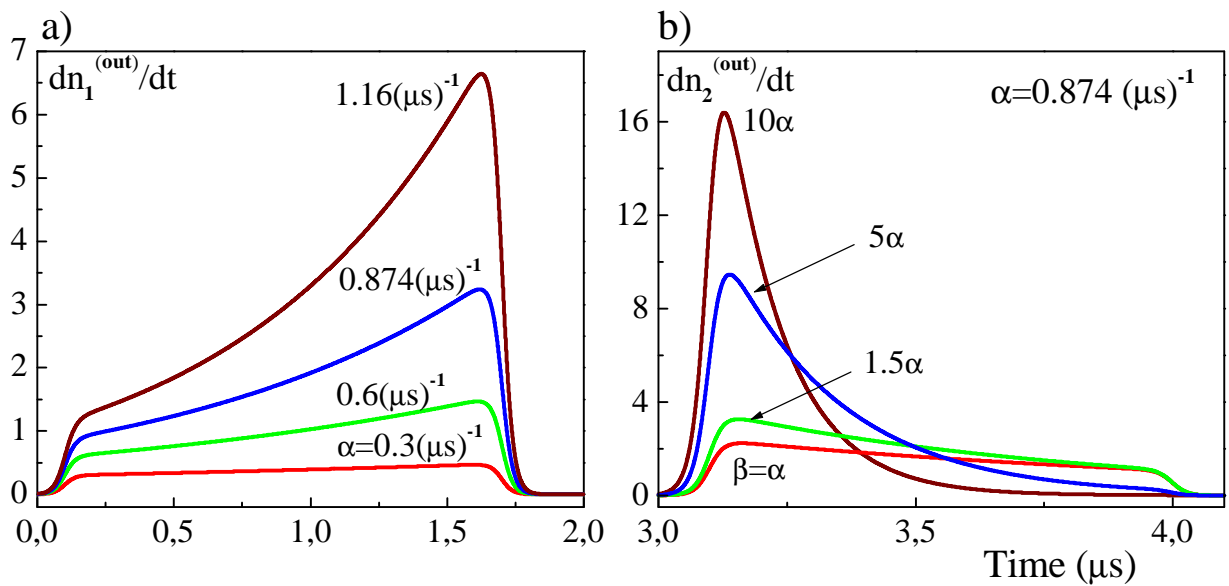
\bigskip Fig.2. a) Stokes photon flux $dn_{1}^{(out)}/dt$ as a function of
time for different values of write pulse intensity. The curves are labeled
with the corresponding values of $\alpha $. The total number of Stokes
photons emitted from the cavity are determined by the areas of the
corresponding peaks. The duration of write pulse T$_{W}=1.6\mu s.$ b)
Anti-Stokes photon flux $dn_{2}^{(out)}/dt$ for fixed number of output
Stokes photons $n_{1}^{(out)}(T_{W})=3$ and different values of read laser
intensity. The read pulse duration T$_{R}=1\mu s.$ The time-delay $\tau
_{d}=1.4\mu s$ and $\gamma _{c}=0.03\ (\mu s)^{-1}.$

\bigskip

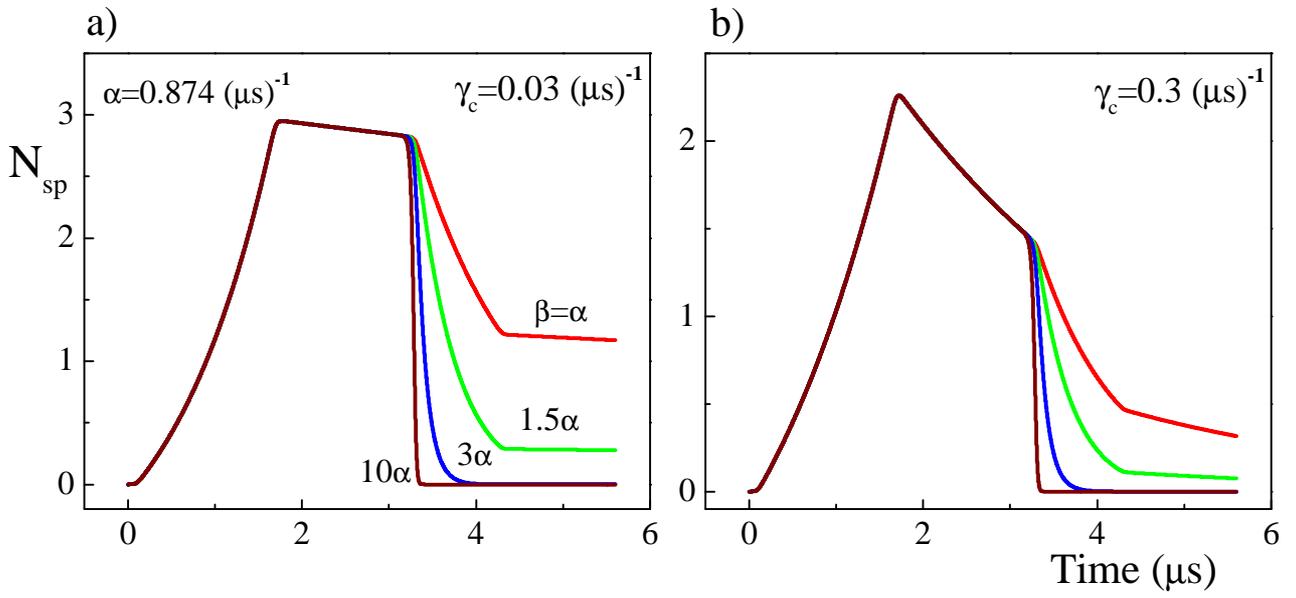
Fig.3. a) Average number of spin-wave excitations N$_{sp}=\left\langle
S^{+}(t)S(t)\right\rangle $ as a function of time for fixed number of
detected Stokes photons $n_{1}^{(out)}(T_{W})=3$ and the values of read
laser intensity, as in Fig.2b. $\gamma _{c}=0.03\ (\mu s)^{-1}.$ b) The
same, but $\gamma _{c}=0.3\ (\mu s)^{-1}$.

\bigskip

\end{document}